\documentclass[useAMS,usenatbib]{mn2e_mod}
\usepackage{graphicx}
\usepackage{amsmath}
\usepackage{amssymb}
\usepackage{deluxetable}

\newcommand{\Msol}{M$_{\odot}$}
\def\fe{\mbox{Fe\hspace{.5pt}{II}}}

\title {An emerging coherent picture of red supergiant supernova explosions}

\author[Poznanski]
{Dovi Poznanski$^{1}$\thanks{dovi@tau.ac.il}\\
$^{1}$School of Physics and Astronomy, Tel-Aviv University, Tel Aviv 69978, Israel.\\}

\begin{document}

\maketitle

\label{firstpage}
\begin{abstract}

Three lines of evidence indicate that in the most common type of core collapse supernovae, the energy deposited in the ejecta by the exploding core is approximately proportional to the progenitor mass cubed. This results stems from an observed uniformity of light curve plateau duration, a correlation between mass and ejecta velocity, and the known correlation between luminosity and velocity. This result ties in  analytical and numerical models together with observations, providing us with clues as to the mechanism via which the explosion of the core deposits a small fraction of its energy into the hurled envelope. 

	%

\end{abstract}



\vspace{1cm}
\section{Introduction}\label{s:intro}

The most common massive stars become red supergiants before ending their lives as Type II-P supernovae (SNe II-P). This fact is now based on numerous progenitor stars that have been identified in archival images, taken before the supernova exploded \citep[see review in][]{smartt09a}. In the most robust cases, post-mortem images have shown that the presumed progenitor had indeed disappeared. SNe II-P are observationally defined by the prominent lines of  hydrogen in their spectra as well as a pronounced plateau phase in their light curve \citep[][]{filippenko97}, both indicative of a massive hydrogen envelope.

Nevertheless, it is quite challenging to connect the more detailed observables of the optical transient to the intrinsic properties of the exploding star. Various analytical models have been put forward and find the dependence of the  supernova luminosity and plateau duration, ${\rm L}$ and  ${\rm t_p}$, on the kinetic energy, ejecta mass, and initial radius, ${\rm E}$, ${\rm M_{ej}}$, and ${\rm R_*}$, respectively \citep{arnett80,woosley88,chugai91,popov93}. A common feature of these various models is that, while they differ on the assumed physics, they arrive at an identical relative contribution of the mass and the energy to the timescale, regardless of the details. Generally speaking, they find that 
\begin{equation}\label{eq:1}
{\rm t_p \propto {\rm E}^{-\alpha}\,{\rm M}^{3\alpha}_{\rm ej}}\,{\rm R_*}^{\beta}.
\end{equation}
The parameters $\alpha$ and $\beta$ depend on the model, but are usually close to $0.15-0.2$. 

With no obvious a-priori relation between the intrinsic parameters that will cancel out their contribution, it is hard to reconcile equation \ref{eq:1} with observational evidence that plateau durations have a narrow distribution centered around 100 days \citep[e.g.,][]{poznanski09,arcavi12}. However, if one posits that the mass is for some reason proportional to the velocity, 
\begin{equation}\label{eq:2}
	{\rm M_{\rm ej}} \propto {\rm v},
\end{equation}
then the energy, ${\rm E} \propto {\rm M_{\rm ej}v^2}$ can be written as
\begin{equation}\label{eq:3}
	{\rm E} \propto {\rm M_{\rm ej}^3}.
\end{equation}
Consequently one gets a constant plateau duration, as long as the dependence of the radius is weak, as predicted by some of the models.

Additionally, it has been strongly established that the luminosity of SNe II-P correlates with the photospheric velocity during the plateau phase, as traced by the Fe II lines \citep{hamuy02}. The correlation of the ejecta velocities of SNe II-P -- as measured from the blueshift of the \fe\ $\lambda5169$ absorption feature -- with their luminosities makes these SNe promising `standardizeable candles' \citep{nugent06,poznanski09,poznanski10}. This is further supported by numerical efforts showing that indeed this line traces the photosphere reasonably well \citep{dessart05a} and obtaining the luminosity-velocity correlation using a grid of models \citep{kasen09}. 
The empirical finding is that the magnitude of Type II-P SNe correlate with photospheric velocity with a parameter of order 5 \citep{hamuy02,poznanski09}, such that $\mathcal{M} = -2.5\, \log({\rm L}) \propto 5\, \log ({\rm v})$, or ${\rm L} \propto {\rm v^2}$. 

Note however, that the luminosity is roughly the available energy internal to the radius of the emitting object over time. Replacing the energy with the energy remaining after adiabatic losses,
\begin{equation}
	{\rm L} \propto \frac{\rm E_{int}({\rm R})}{\rm t} \propto {\rm E}\,\frac{\rm R_*}{\rm R(t)\,t}.
\end{equation}
Further replacing ${\rm R(t)}$ with ${\rm vt}$, and ${\rm E}$ with ${\rm M_{\rm ej}v^2}$ the luminosity scales as 
\begin{equation}
	{\rm L} \propto \frac{\rm M_{\rm ej}vR_*}{\rm t^2},
\end{equation}
where ${\rm t}$ is the typical timescale, which can be replaced with ${\rm t_p}$. A constant plateau length and ${\rm L} \propto {\rm v^2}$
therefore imply again that ${\rm M_{\rm ej}\propto v}$, when assuming only small variations in radius. 

I consequently find that the uniform plateau duration and the luminosity-velocity relation are mutually consistent and come naturally from the analytical and numerical models (see more in Section \ref{s:models}) if one posits that the kinetic energy deposited in the ejecta is somehow proportional to the cube of the ejected mass, which equally implies a linear relation between mass and velocity.




\begin{figure*}\label{f:1}
\includegraphics[width=0.85\textwidth]{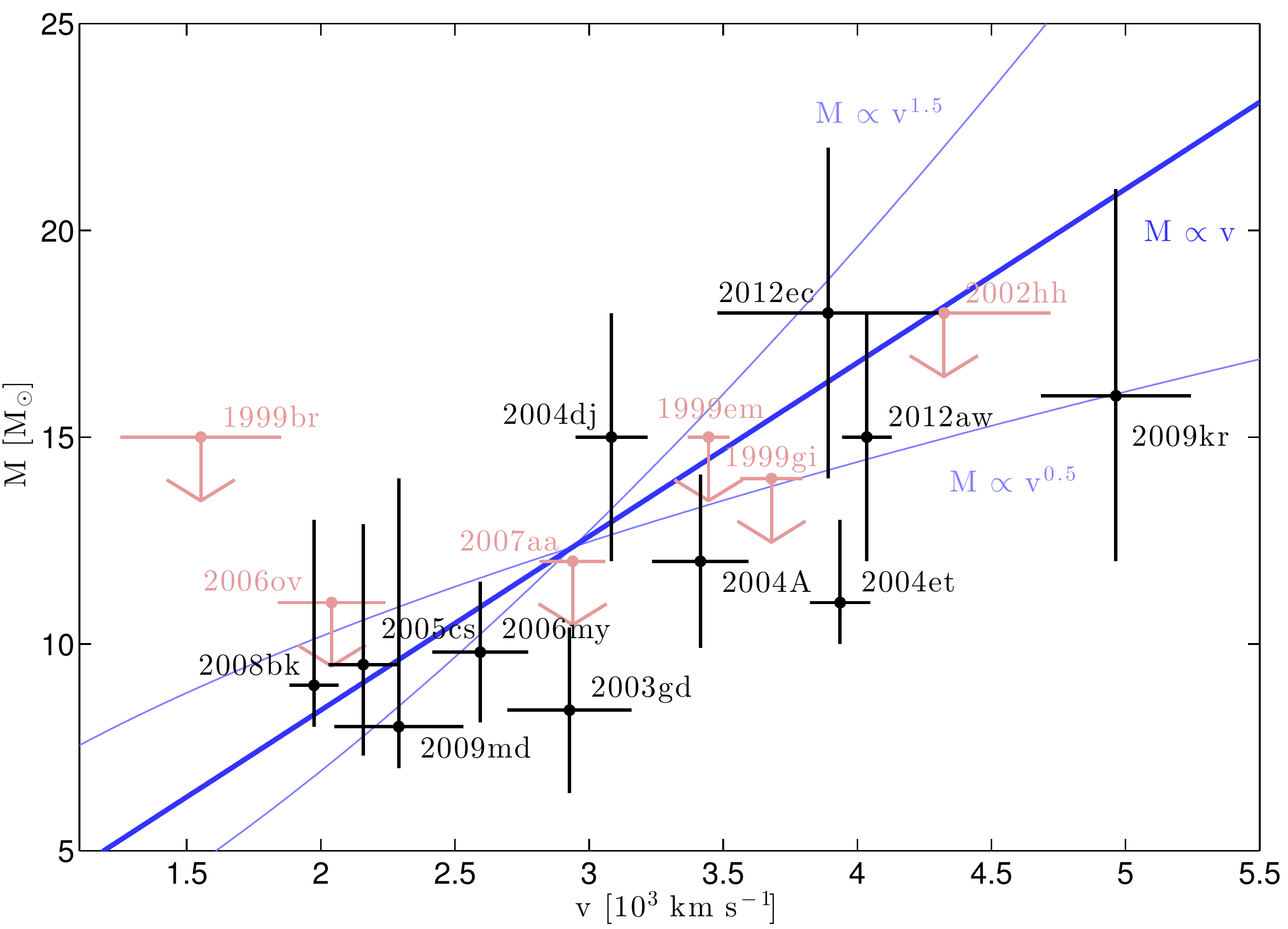}
\caption{Black crosses show the ejecta velocities and initial stellar masses for a sample of Type II-P supernovae with progenitors detected in pre-explosion images. Objects with only upper limits on the mass appear as pink arrows. Data are consistent with a linear trend between velocity and mass, as can be seen with the best fit power law in dark blue. For comparison, steeper and shallower dependences are plotted in light blue.}
\end{figure*}

I have tested this conjecture by compiling from the literature a comprehensive set of initial stellar masses (or mass limits) for 23 Type II-P supernovae \citep{smartt09,fraser11,crockett11,maund13,van-dyk12,fraser12,jerkstrand12,maund13a}. For 17 of these supernovae, I can derive ejecta velocities. As I show below, I find that the progenitor masses and velocities are indeed consistent with a linear relation. A combination of these lines of evidence -- plateau duration, luminosity-velocity relation, mass-velocity relation -- all point towards a puzzling intrinsic correlation between kinetic energy and initial mass, such that ${\rm E} \propto {\rm M^3}$. In the following sections I present the sample, the new found mass-velocity relation, and discuss the implications of these findings.

\section{sample}

\subsection{Masses}

The focus of this study is SNe II-P. Determining the mass of their progenitors depends on an array of modeling tools and assumptions, most notably regarding the dust that enshrouds them, the distances to their host galaxies, and the details of their pre-SN evolution. These difficulties are unavoidable in attempting to link a handful of observed colours and brightnesses, measured a cosmic moment before explosion at a post main-sequence stage, to initial conditions set millions of years before. This entails large statistical uncertainties, and equally substantial systematic ones. As this field progresses, methods improve, further observations are obtained after the SN has faded, and some determinations are revised. 

\citet{smartt09} and later \citet{fraser11} compiled a comprehensive list of progenitor mass determinations, including a self consistent re-derivation of some of the masses. \citet{crockett11} revisit three of these masses, \citet{maund13} revisit five more, 
and I further add SNe 2009kr \citep{elias-rosa10,fraser10}, and the recently discovered 2012aw \citep{van-dyk12,fraser12,kochanek12} and 2012ec \citep{maund13a}. In table 1 I list the masses as compiled from these sources. 

\citet{maund13} find that the previously identified progenitor of SN\,1999ev is actually still present in images taken after the SN has faded. They discuss various possibilities. First, the unresolved source may be a cluster that is now one star short. Second, the source may be unrelated. Last, the new source might be a light echo, and the progenitor has indeed disappeared. The authors find the probability for the last two options to be rather low, and they therefore discard the previously measured mass for this progenitor. I follow suit and discuss this SN further in Section \ref{s:indiv}. 


The progenitor mass of SN\,2004et is difficult to measure, since it is based on ground based data and requires non-trivial subtraction of neighboring stars that are contaminating the photometry. The $8^{+5}_{-1}$\,\Msol\ mass derived by \citet{crockett11} is being revised by Fraser et al. (in prep) as quoted by  \citet{jerkstrand12}. The new higher mass of $11^{+2}_{-1}$\,\Msol\ is also consistent with competing efforts (Van Dyk, private communication). For this reason I use the higher mass for SN\,2004et as quoted in \citet{jerkstrand12}. 


There is some risk in including SN\,2009kr in my analysis. As discussed below it is not a `classical' II-P and may actually come from a different progenitor channel. There is also some controversy regarding its mass. While \citet{elias-rosa10} advocate a mass of $21\pm3$\,\Msol, \citet{fraser10} argue for a mass of $15^{+5}_{-4}$\,\Msol, claiming that the comparison to stellar tracks that have not finished core helium burning is misguided. One should further note that the progenitor is consistent with recent stellar tracks for 20\,\Msol\ stars that include the effects of pulsation-driven superwinds \citep{yoon10}. The luminosity and temperature measured for the progenitor are in perfect agreement with their pre-SN predictions. Clearly, for Type II SNe of these higher masses, the difficulty to model the effects of mass loss make mass determinations even less robust than for the run-of-the-mill Type II-P SNe. In practice, all these masses are statistically consistent with each other, and while I use the mass derived by \citet{fraser10} (adjusted for solar metallically; Smartt private communication) either mass would be consistent with my results below.


\subsection{Supernovae II-L}
The historic division between Type II-P and II-L SNe is photometric, and is based on the observation that some Type II SNe do not show a pronounced plateau \citep[e.g.,][]{barbon79,poznanski02}. However this segregation is based on observations in blue colours, where most SNe show some decline, and predates the spectroscopic discovery and definition of type IIn and IIb SNe that indeed often show a faster-declining light curve. The first are mostly powered by interaction with the circumstellar medium, and can therefore have wildly different light curves depending on the mass-loss history. The latter are intermediate between Type I and Type II and are interpreted as being transition objects with a thin hydrogen shell, visible in early spectra until the photosphere moves more deeply in. Their photometric evolution is therefore more similar to the $^{56}$Ni powered Type Ib/c SNe. 

SN\,1979C, and SN\,1980K are often referred to as the prototypical II-L SNe, but in the same breath are called `over luminous' and `peculiar' \citep{filippenko97}. When studying the cosmological utility of SNe II-P it has been noted that some of the otherwise normal looking SNe which show a rather subtle decline in $I$-band photometry, are outliers on the Hubble diagram \citep{poznanski09}. This has led some to classify objects that decline by about 1\,mag over the $\sim100$\,day plateau as II-L \citep[e.g.,][]{li11a}. 

Recently, studying a sample of $R$-band light curves \citet{arcavi12} have found that indeed there is a subsample of Type II SNe that decline by about that much, and that they do not form a continuum with the normal plateau objects. Studies of the spectroscopic features of these SNe are still preliminary \citep{schlegel96}. At least from a photometric perspective SN\,2009kr seems to belong to this class. Whether or not these indeed form a distinct class of SNe, in terms of their photometric and spectroscopic, as well as progenitor properties, is still a mostly open question.

\begin{deluxetable}{lcl}\label{t:1}
\tablewidth{0pt}
\tabletypesize{\scriptsize}
\centering
\tablehead{
\colhead{SN name} &
\colhead{M (M$_{\odot}$)} &
\colhead{Source}}
\tablecaption{Masses}
\startdata
1999an & $<\,18$ & \cite{smartt09} \\ 
1999br & $<\,15$ & \cite{smartt09} \\ 
1999em & $<\,15$ & \cite{smartt09} \\ 
1999gi & $<\,14$ & \cite{smartt09} \\ 
2001du & $<\,15$ & \cite{smartt09} \\ 
2002hh & $<\,18$ & \cite{smartt09} \\ 
2003gd & $ 8^{+2}_{-2}$ & \cite{maund13} \\ 
2003ie & $<\,25$ & \cite{smartt09} \\ 
2004A & $12^{+2}_{-2}$ & \cite{maund13} \\ 
2004am & $12^{+7}_{-3}$ & \cite{smartt09} \\ 
2004dg & $<\,12$ & \cite{smartt09} \\ 
2004dj & $15^{+3}_{-3}$ & \cite{smartt09} \\ 
2004et & $11^{+2}_{-1}$ & \cite{jerkstrand12} \\ 
2005cs & $10^{+3}_{-2}$ & \cite{maund13} \\ 
2006bc & $<\,12$ & \cite{smartt09} \\ 
2006my & $10^{+2}_{-2}$ & \cite{maund13} \\ 
2006ov & $<\,11$ & \cite{smartt09} \\ 
2007aa & $<\,12$ & \cite{smartt09} \\ 
2008bk & $ 9^{+4}_{-1}$ & \cite{smartt09} \\ 
2009kr & $16^{+5}_{-4}$ & \cite{fraser10}\tablenotemark{a}\\ 
2009md & $ 8^{+6}_{-1}$ & \cite{fraser11} \\ 
2012aw & $15^{+3}_{-3}$ & \cite{kochanek12}\tablenotemark{b}\\ 
2012ec & $18^{+4}_{-4}$ & \cite{maund13a} \\ 
\enddata
\tablenotetext{a}{Adjusted to solar metallically (S. Smartt, private communication).}
\tablenotetext{b}{Using the luminosity derived by \citet{kochanek12} and the STARS stellar tracks (S. Smartt, private communication).}

\end{deluxetable}

\subsection{Velocities}

I compile spectra taken during the plateau phase using publicly available resources \citep{yaron12}, supplemented mostly by published spectra obtained by the group of A. Filippenko at UC Berkeley \citep{silverman12}. Velocities are measured using the same algorithm used for the standardized candle method \citep{poznanski09}. Briefly, using the SNID code \citep{blondin07}, every spectrum is cross-correlated with a set of high signal-to-noise spectra with manually measured velocities.  The resulting velocities are propagated to day 50 using the power law $v(50)=v(t)(t/50)^{0.464\pm0.017}$ \citep{nugent06} and averaged using their uncertainties as weights. I keep only objects for which the resulting velocity uncertainty is smaller than 1000\,km\,s$^{-1}$. 

Since SNe II-P evolve slowly during the plateau phase, both spectroscopically and photometrically, their properties in the middle of the plateau are less sensitive to systematics arising from uncertainties regarding the precise explosion date. As the velocity approximately declines like $t^{-0.5}$, an uncertainty of $\pm10$ days implies a velocity uncertainty of only $\pm10$ percent. 

Still, there can be a significant uncertainty for some SNe that were discovered late. In order not to bias velocity measurements, I do not use spectroscopic information for phase determination, relying instead on photometric constraints alone. For most SNe the explosion date is conservatively set as the midpoint between discovery and last non-detection with an uncertainty that spans that time range. For some objects, where such a constraint is weak and overestimates the uncertainty, I use the end of the plateau phase as an indicator that $100\pm10$ days have passed since explosion. Table 2 summarizes the spectroscopic data.

\section{Velocity vs. Mass}

It has been shown that sub-luminous SNe II-P tend to have progenitor masses at the low end of possible masses, i.e., near 8\,\Msol\ \citep{fraser11}, but it is unclear whether there is indeed a correlation between mass and luminosity that extends beyond the lower-mass regime, or if there are two populations. A major hurdle is that SN luminosities are difficult to measure. Most importantly, the luminosity depends on the amount of dust and the distance that one measures or assumes. However velocities are less prone to systematic uncertainties. 

Figure \ref{f:1} shows the correlation between velocity and mass for the sample. With the large uncertainties in mass, their non-gaussian distribution, and the non negligible uncertainties in velocity, standard out-of-the-box tools often used by astronomers will fail to give a reliable estimate of the significance of this result. In order to make it a linear (and symmetric) fit, in the discussion below I take the logarithm of both velocity and mass, and subtract the sample mean from each variable. The slope of the best fit line is the best fit power. 

There are multiple solutions in the literature for dealing with   uncertainties on both variables, and a lack of straightforward definition of the dependent or independent variable \citep{babu92,akritas96,tremaine02}. A simple example can elucidate the difficulty. If we were to fit a linear model to random uncorrelated $x$ and $y$ variables, a standard least square fit will find a best fit slope for $(y|x)$ of zero - parallel to the abscissa. However, inverting the variables and fitting for the slope of $(x|y)$, one would get zero as well, parallel to our previous ordinate. Of course neither are significant, but this example shows how when the scatter is large the choice of dependent or independent variable can dramatically change the result. One simple solution, called the bisector least-square method, is to take the line that bisects the two solutions presented above. Another method, called the orthogonal least square method, is to measure orthogonal distances from the best fit line rather than vertical distances (i.e., along the $y$-axis). This creates the needed symmetry between the variables as well. 

In order to include the mass limits and treat them as measurements, I must assume some prior distribution on their mass, as well as an upper error which is not commonly reported for limits (i.e., if an object has a limit of M\,$<10$\,\Msol\ could the progenitor have with some probability a mass of 11\Msol\ or 10.1\Msol?). I assume upper errors that are a third of the typical error on mass for all the mass determinations (a crude approximation to the fact that non-detections are typically $3\sigma$) and lower errors that reach to a minimum mass of 6\Msol. 

I generate thousands of simulated samples which allows me to measure the uncertainty in the slope. For each of the thousands of instances, 17 masses and matching velocities are randomly drawn from normal distributions having the measured masses and velocities as means, and their uncertainties as standard deviations. By bootstrapping, i.e., selecting 17 random SNe from the sample (with repetition), I can ascertain that no single SN is driving the relation. 

I perform a Pearson correlation test on the samples, and obtain a correlation coefficient distribution which is centered on $0.42\pm0.2$. The same test on randomly permuted data-pairs returns $-0.01\pm0.2$, with values higher than 0.42 for only 4 percent of the samples. A Spearman test gives the same result. Therefore the data favor a correlation at the 96 percent level.  

Using the bisector method I find a slope of $0.95\pm0.15$. An orthogonal distances algorithm finds slopes of about $0.8\pm0.3$. Various other tests, correlation measures, resampling methods, all give a similar conclusion, that the correlation is significant and its slope of order unity. 

One should also ask why many of the SNe with mass limits lie so close to the  correlation. First, it should be noted that I did not attempt to collect or measure velocities for SNe with very weak progenitor mass limits (beyond the few I had available). I therefore do not have many objects with mass limits lying in the high-mass sector of the figure (though this would add 2 SNe at most). Second, for volumetric reasons most objects found by surveys are near the flux limit. If indeed there is a correlation between mass and velocity and between velocity and luminosity, given a non-detection of the progenitor, the posterior mass probability of every SN is skewed towards higher mass. If there is no correlation there should be no such preference, and the limits should not lie so close to the relation. Had there been no correlation one would have expected the limits to populate the parameter space much more uniformly. The mass limits by themselves therefore suggest a correlation. 

The best fit trend has a $\chi^2$ smaller than the number of degrees of freedom, which is expected considering the large error bars. This also indicates that current data are unable to constrain the intrinsic scatter in the mass-velocity relation.

\subsection{Individual progenitors}\label{s:indiv}

While the relation between mass and velocity shows as a significant scatter, either intrinsic or driven by the large uncertainties in the progenitor masses, there are a few tentative conclusions regarding individual progenitors that one could draw by assuming that indeed the relation holds. 

First, I find a very low velocity for SN\,1999ev of about 2200\,km\,s$^{-1}$, which would make a $16^{+6}_{-4}$\,\Msol\ progenitor a significant outlier to the mass-velocity relation. The low velocity reinforces the findings of \citet{maund13} and points to a low mass progenitor, near 8\,\Msol. This example indicates that the mass-velocity relation could sometimes help us differentiate between progenitor candidates in ambiguous cases. 

For SN\,2004et as well, the correlation here can be seen as an indication that the earlier mass estimate was perhaps somewhat underestimated, though it is only about $2\sigma$ away from the best fitting relation. 

As mentioned above, there is some controversy regarding the mass of SN\,2009kr. The relation seems to point towards a mass in the 20\Msol\ range rather than the 15\Msol\ range, which I have used here.

\begin{figure}
\includegraphics[width=3.25in]{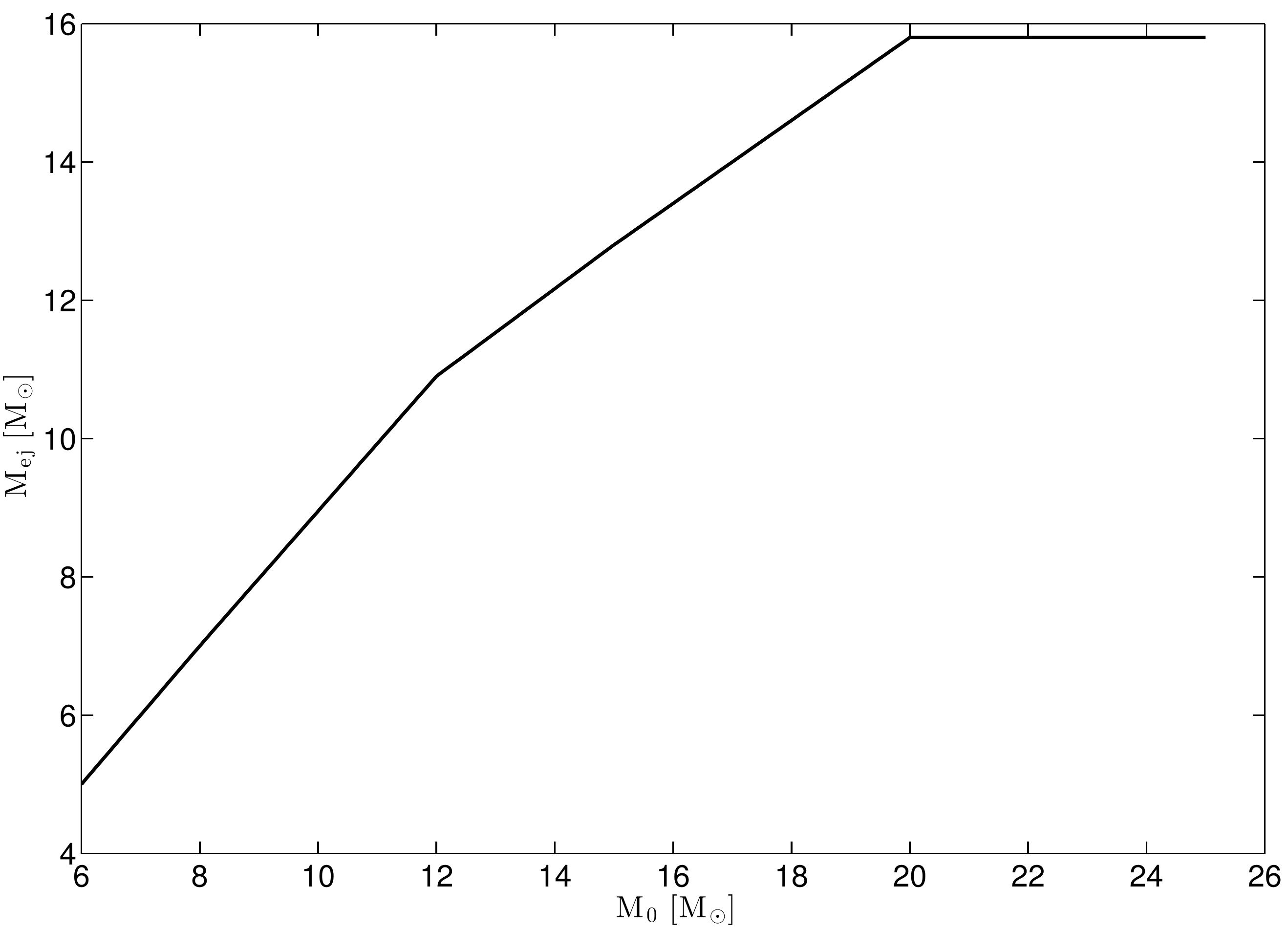}
	\caption{Relation between initial mass and ejecta mass, based on the models presented by \citet{kasen09}. Since for most masses the relation is linear, and the uncertainties are rather large, the two masses are virtually indistinguishable with regards to the mass-velocity relation.}\label{f:2}
\end{figure}

\begin{figure}
\includegraphics[width=3.25in]{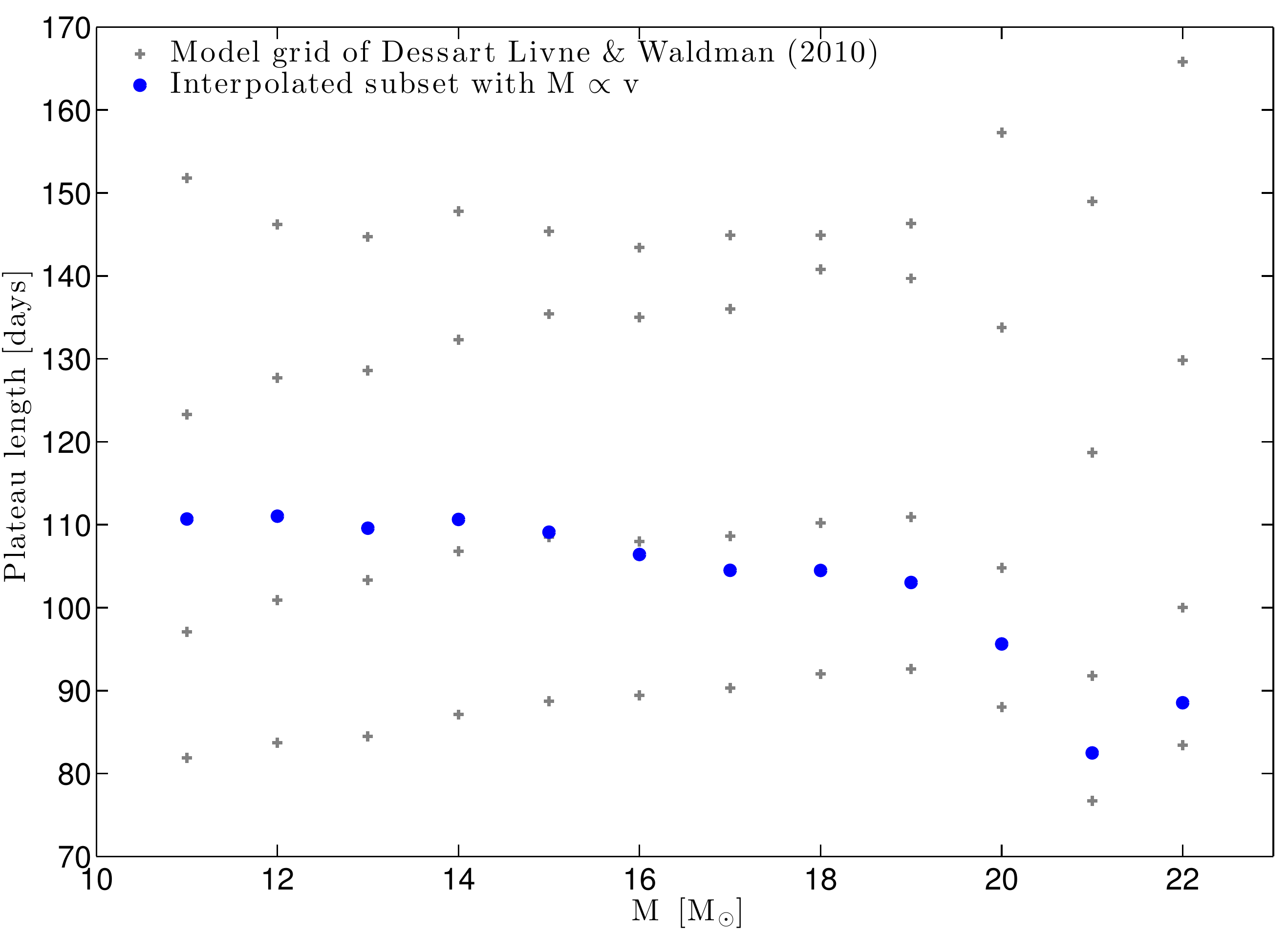}
\caption{Masses and plateau lengths for a grid of models \citep{dessart10a} in grey crosses, and for an interpolated subset that follows ${\rm M}\propto {\rm v}$ in blue dots. The wide range of a-priori possible timescales is substantially reduced as predicted from observations and the scaling relations. These results further predict a slight decrease in duration with mass.}\label{f:3}
\end{figure}

\begin{figure}
\includegraphics[width=3.25in]{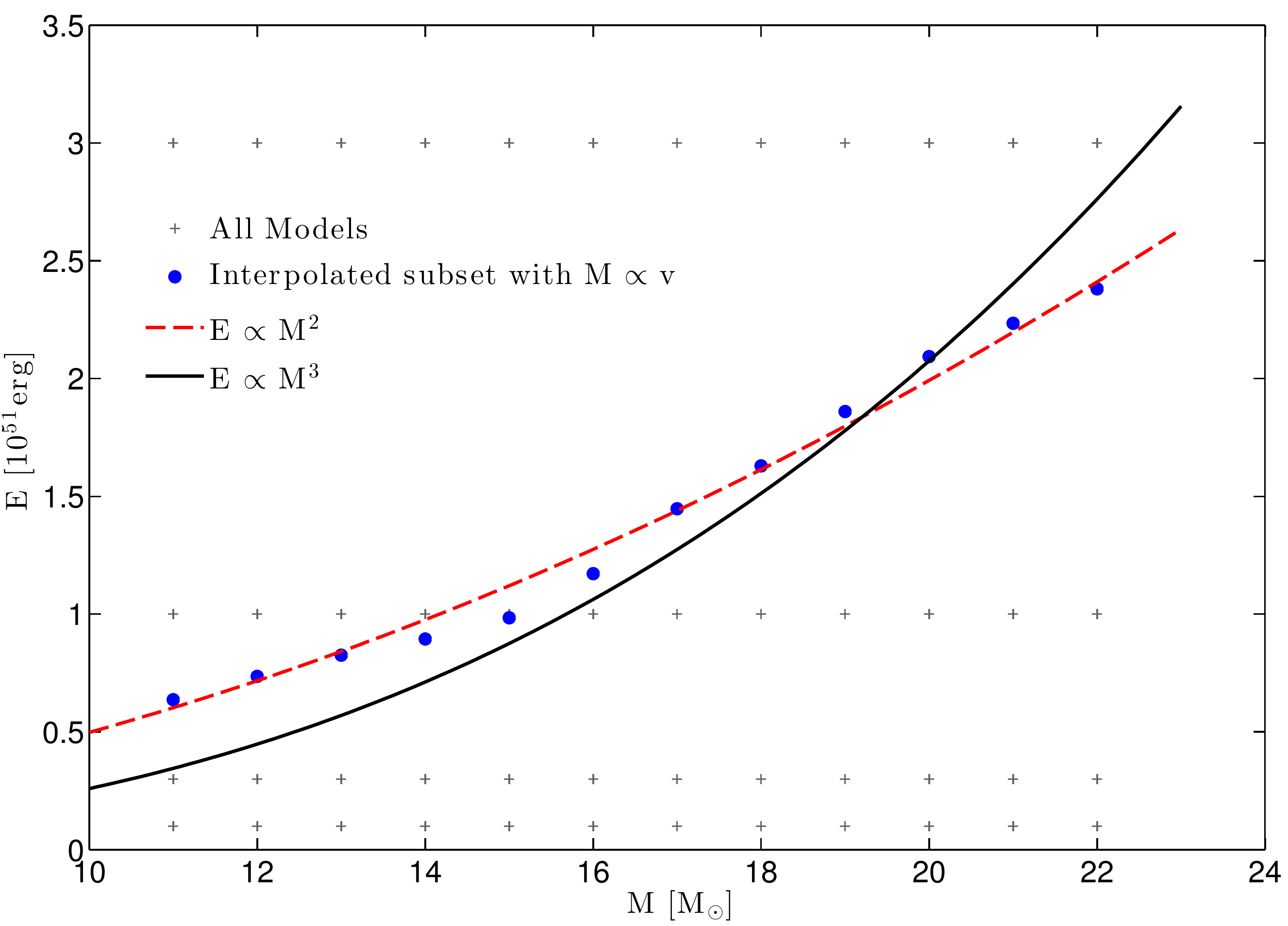}
\caption{
Kinetic energy vs. initial mass, interpolating the models in \citep{dessart10a} that obey ${\rm M}\propto {\rm v}$. This confirms the finding that energy rises with mass, albeit with a somewhat smaller power. See however text for a possible explanation, as well as Figure 5.}\label{f:4}
\end{figure}

\begin{figure}
\includegraphics[width=3.25in]{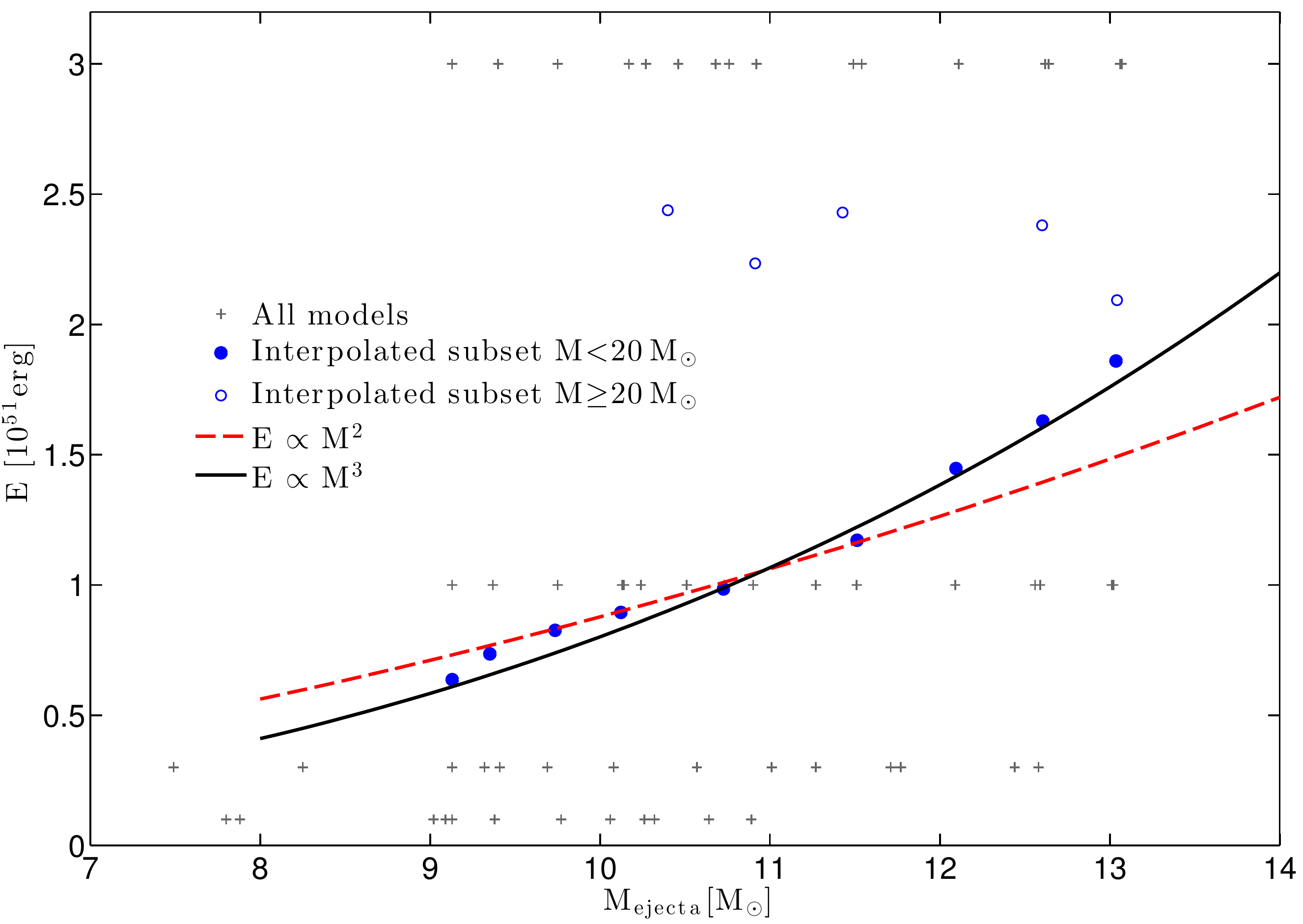}
\caption{
Same as Figure 4, replacing initial masses with ejecta masses. The correlation now is closer to ${\rm E}\propto {\rm M^3}$, but breaks at large mass, as mass loss becomes highly non-linear.}\label{f:5}
\end{figure}

\section{Initial vs. Ejected Mass and Comparison to Numerical Models}\label{s:models}

There are various masses involved -- the initial mass, the pre-explosion mass, the remnant mass, and the ejecta mass. I eventually treat the initial and ejecta mass as practically interchangeable. The difference between the two masses are the mass of the stellar remnant, for which it is safe to assume a mass of order 1--2\,\Msol (hence a systematic variance much smaller than the uncertainties in the progenitor masses), and the effects of mass loss from stellar birth to demise. The latter process is poorly understood, but it is likely that for relatively small masses (i.e., 6\,\Msol$<\, {\rm M}\, <$15\,\Msol) there is an approximately linear dependence of the mass loss on mass. At higher masses mass loss becomes more efficient and can effectively leave a smaller pre-SN star despite the larger initial mass. This picture is consistent with the models in Table 2 of \citet{kasen09}. 

Using this table as a prescription (see Figure \ref{f:2}) to transform the initial masses of the progenitors to ejected masses does not alter the picture in any significant way. The transformation is essentially linear for all SNe and does not introduce a significant change, within the existing mass uncertainties. Because this change does lower higher-mass objects more significantly, the best fit power-law, while still being consistent with a linear relation, is a few percent smaller for both analysis methods presented above. 

A grid of models has been calculated \citep{dessart10a} where the authors predominantly vary the mass between 11 and 30\,\Msol\ and the kinetic energy that is pumped into the ejecta, probing the range between 0.1 and 3 times 10$^{51}$\,erg. As shown in Figure \ref{f:3}, when applying the constraint ${\rm M}\propto {\rm v}$ the wide range of plateau lengths that the models predict are narrowed down to a very tight distribution. 
In addition, Figure 4 shows that the energy values that follow this same constraint also correlate with mass, with a best fit power of ${\rm E}\propto {\rm M^2}$. This value is somewhat lower than found from observations, but could be understood as resulting from the conversion between initial and ejecta mass. As shown in Figure 5, replacing initial masses with ejecta masses from the models one recovers the ${\rm E}\propto {\rm M^3}$ relation. This however breaks down, as initial masses exceed 20\,\Msol, at which point mass loss can bring a more massive star to have smaller ejecta, linking the realm of II-P SNe to their more massive II-L cousins, of which SN\,2009kr is likely an example. 

\section{discussion}

Had there been a roughly constant energy deposited by the explosion, ${\rm E}\propto{\rm Mv}^2 = {\rm const}$, the velocity would appear to decline with increasing mass, as ${\rm v}\propto{\rm M}^{-1/2}$. However, since I find indications for an approximately linear relation ${\rm M}\propto{\rm v}$, then the energy is far from constant and actually depends strongly on the mass, such that ${\rm E}\propto{\rm M^3}$.

Using this relation, the dependences on energy and mass in equation 1 cancel out, and the plateau duration, ${\rm t_P}$, depends only on the radius to a small power (or not at all, depending on the model). A variation of nearly a factor of 2 in radii will translate at most to a 15\% variation in plateau length, which is consistent with observations. Radii measured for many of the progenitors in this sample seem to broadly agree with this picture \citep{smartt09}. This is further consistent with the relation between luminosity and velocity. Ongoing surveys that probe the supernova evolution during the first few days past explosion can constrain the radius at the time of explosion \citep{gal-yam11,sagiv13} and further test these results.



Additionally, from the numerical models of \citet{dessart10a}, I find that the plateau duration is somewhat anti-correlated with mass and energy, such that higher-mass models have shorter plateaus. This is consistent with the current picture in which the highest mass type II supernovae do not have a plateau at all, having lost most of their envelope via winds. There are some early indications that this prediction actually holds (Faran et al. in prep.). 

I therefore find multiple independent pieces of evidence for a strong dependence of the explosion energy on the initial mass, of order ${\rm E} \propto {\rm M^3}$. The implications of this puzzling result are not obviously clear. Nevertheless, it is likely an important clue regarding the mechanism that couples the large energy released during core collapse to the ejected mass, and leading to a successful explosion. A factor of about 2.5 in progenitor mass ($\sim8-20$\,\Msol) translates into a factor of 15 in energy. 

It was originally posited that as the star's core reaches iron and collapses,  core bounce generates the shock that disrupts the star. However, it was later argued that photodisintegration and neutrino losses will cause the shock to stall, requiring more complicated mechanisms to revive the shock and produce a successful SN (see \citealt{woosley86} for a thorough review). These arguments, while robust from a theoretical standpoint, can now be considered to have observational evidence. It would be difficult to explain a strong dependence of the energy on mass, as stars of different masses (up to about 20\,\Msol at least) have quite similar cores (e.g., \citealt{woosley95,fryer12}). Therefore, the most natural place to allow for such a variation is not in the total available energy, but in the efficiency of the deposition process, be it via neutrinos, instabilities, or any other mechanism summoned to successfully explode the star (e.g., see \citealt{ugliano12}). 

If the energy deposition process is slow \citep{dessart10b}, the binding energy of the envelope has an important impact on the energetics of its subsequent expulsion. However, at best the binding energy scales as ${\rm M}^2$, or even linearly with mass (see figure 5 of \citealt{woosley95}). 
An additional effect of the binding energy is via an observational bias, as a core collapse event with insufficient energy to unbind the mantle will fail to produce a SN. This argument can explain at least the paucity of events with low energy and large mass. A more thorough interpretation of ${\rm E}\propto{\rm M^3}$ relation and its implications will await detailed theoretical analysis.

\section*{Acknowledgments}
First and foremost, I am grateful to my friends and colleagues who have systematically gathered the large body of observations that I have used here. 
	I am no less thankful to 
	A. Gal-Yam,
	R. Helled,
	I. Kleiser, 
	A. Levinson,
	E. Livne,
	A. Loeb,
	D. Maoz,
	E. Nakar, 
	H. Netzer,
	B. Trakhtenbrot, 
	and S. Van Dyk, for valuable advice regarding many different aspects of this work. The same gratitude goes towards the very careful referee S. Smartt who helped improve my work. I am further indebted to A. Filippenko for allowing me access to his impressive database of supernova data. S. Benetti and the Padova supernova group are thanked for the spectrum of SN\,2004dg. M Stritzinger and the Carnegie supernova program are thanked for the spectra of SN\,2008bk. A. Gal-Yam and the Palomar Transient Factory are thanked for the spectra of SN\,2012aw. Based on data products from observations made with ESO Telescopes at the La Silla Paranal Observatory under programme 188.D-3003: PESSTO (the Public ESO Spectroscopic Survey for Transient Objects). This work made use of the Weizmann interactive supernova data repository  (\texttt{www.weizmann.ac.il/astrophysics/wiserep}), as well as the NASA/IPAC Extragalactic Database (NED) which is operated by the Jet Propulsion Laboratory, California Institute of Technology, under contract with the National Aeronautics and Space Administration. I further acknowledge the support of the Alon fellowship for outstanding young researchers, from the Raymond and Beverly Sackler Chair for young scientists, and the Dark Cosmology Center which is funded by the Danish National Research Foundation for hosting me while working on this topic.

\bibliographystyle{mn2e} 
\bibliography{myBIBTeX}

\clearpage
\begin{deluxetable}{lccccl}\label{t:2}
\tablewidth{0pt}
\tabletypesize{\scriptsize}
\tablecaption{Velocities}
\tablehead{
\colhead{SN name} &
\colhead{cz$_{\rm CMB}$ (km\,s$^{-1}$)} &
\colhead{Explosion date} &
\colhead{v$_{\rm Fe\,II}$(50d) (km\,s$^{-1}$)} &
\colhead{Number of spectra} &
\colhead{Source} }
\startdata
1999br &  960 & -- & $1550\pm300$ & -- & \cite{hamuy02} \\ 
1999em &  717 & 25/10/1999\,$\pm  4$ & $3450\pm 80$ & 28 & \cite{leonard02em} \\ 
1999gi &  592 &  7/12/1999\,$\pm  4$ & $3680\pm120$ & 15 & \cite{leonard02a} \\ 
2002hh &   40 & 29/10/2002\,$\pm  2$ & $4320\pm400$ & 6 & \cite{pozzo06} \\ 
2003gd &  657 & 18/ 3/2003\,$\pm 21$ & $2930\pm230$ & 4 & \cite{van-dyk03} \\ 
2004A &  852 &  6/ 1/2004\,$\pm  3$ & $3410\pm180$ & 1 & \cite{hendry06} \\ 
2004dj &  133 & 30/ 6/2004\,$\pm 10$ & $3080\pm130$ & 5 & \cite{leonard06} \\ 
2004et &   40 & 23/ 9/2004\,$\pm  0$ & $3940\pm110$ & 12 & \cite{li05} \\ 
2005cs &  463 & 28/ 6/2005\,$\pm  1$ & $2160\pm130$ & 7 & \cite{li06} \\ 
2006my &  788 & 28/ 8/2006\,$\pm 10$ & $2590\pm180$ & 3 & \cite{li07,chornock10} \\ 
2006ov & 1566 &  7/ 9/2006\,$\pm 10$ & $2040\pm200$ & 6 & \cite{li07,chornock10} \\ 
2007aa & 1465 & 19/ 1/2007\,$\pm 10$ & $2940\pm120$ & 4 & \cite{chornock10} \\ 
2008bk &  230 & 25/ 3/2008\,$\pm  2$ & $1970\pm 90$ & 2 & \cite{van-dyk12a} \\ 
2009kr & 1939 & 28/10/2009\,$\pm 10$ & $4960\pm280$ & 8 & \cite{elias-rosa10} \\ 
2009md & 1308 & 27/11/2009\,$\pm  8$ & $2290\pm240$ & 4 & \cite{steele09} \\ 
2012aw &  778 & 16/ 3/2012\,$\pm  1$ & $4040\pm 90$ & 9 & \cite{van-dyk12} \\ 
2012ec & 1408 &  7/ 8/2012\,$\pm  5$ & $3890\pm410$ & 4 & \cite{maund13a} \\ 
\enddata
\end{deluxetable}

\end{document}